\documentclass[aps,prc,preprint,groupedaddress,showpacs]{revtex4}
\usepackage{epsfig}
\usepackage{amsmath}

\begin{document}

\preprint{OSU/201-11Li208Pb}

\title{The interaction of $^{11}$Li with $^{208}$Pb}

\author{A.M. Vinodkumar}
\altaffiliation{Dept. of Physics,University of Calicut, Kerala 673635, India}
\author{W. Loveland}
\author{R. Yanez}
\author{M. Leonard}
\author{L. Yao}
\affiliation{Department of Chemistry, Oregon State University,
 Corvallis, OR, 97331.}
\author{P. Bricault}
\author{M. Dombsky}
\author{P. Kunz}
\author{J. Lassen}
\author{A. C. Morton}
\author{D. Ottewell}
\author{D. Preddy}
\author{M. Trinczek}
\affiliation{TRIUMF, Vancouver, British Columbia V6T 2A3 Canada}

\date{\today}

\begin{abstract}

{\bf Background:} $^{11}$Li is one of the most studied halo nuclei.  The fusion of $^{11}$Li with $^{208}$Pb has been the subject of a number of theoretical studies with widely differing predictions, ranging over four orders of magnitude,  for the fusion excitation function.

{\bf Purpose:} To measure the  excitation function for the $^{11}$Li + $^{208}$Pb reaction.

{\bf Methods:} A stacked foil/degrader assembly of $^{208}$Pb targets was irradiated with a $^{11}$Li beam producing center of target  beam energies from above barrier to near barrier energies ( 40 to 29 MeV).  The intensity of the $^{11}$Li beam (chopped) was 1250 p/s and the beam on-target time was 34 hours. The $\alpha$-decay of the stopped EVRs was detected in an $\alpha$-detector array at each beam energy in the beam-off period (the beam was on for $\le$  5 ns and then off for 170 ns). 

{\bf Results:}  The observed nuclidic yields of $^{212/215}$At and $^{214}$At are consistent with being produced in the complete fusion of $^{11}$Li with $^{208}$Pb.  The observed yields of $^{213}$At appear to be the result of the breakup of $^{11}$Li into $^{9}$Li + 2n, with the $^{9}$Li fusing with $^{208}$Pb.  The magnitudes of the total fusion cross sections are substantially less than most theoretical predictions.

{\bf Conclusions:} It is possible to measure the evaporation residue production cross-sections resulting from the interaction of  $^{11}$Li with $^{208}$Pb using current generation radioactive beam facilities.  Both complete fusion and breakup fusion processes occur in the interaction of $^{11}$Li with $^{208}$Pb.  An important breakup process leads to the fusion of the $^{9}$Li fragment with $^{208}$Pb .

\end{abstract}


\maketitle

\section{Introduction}

One of the most active areas of research with radioactive beams is the study of the fusion of weakly bound nuclei, such as the halo nuclei.  The central issue is whether the fusion cross section will be enhanced due to the large size of the halo nucleus or whether fusion-limiting breakup of the projectile will lead to a decreased fusion cross section.

$^{11}$Li is one of the most studied halo nuclei.  $^{11}$Li is a two neutron halo nucleus with S$_{2n}$ = 0.30 MeV.  The fusion of $^{11}$Li with $^{208}$Pb has been the subject of a number of theoretical studies resulting in widely varying predictions for the fusion excitation function.  Figure 1 (taken from the review articles \cite{sig1,sig2} by Signorini) shows the range of predictions of the fusion excitation function.  The most striking feature of Fig. 1 is that the predicted cross sections differ by up to four orders of magnitude. (In the figure legend, the terms ``soft E1", ``BU" and ``1 D Tunn" refer to calculations involving the excitation of the soft dipole mode, breakup of the projectile and a simple one dimensional tunneling, respectively.  Canto refers to \cite{canto}, Takigawa \cite{taki}, Hussein \cite{huss} and Dasso \cite{dasso}.) All calculations involve possible excitation of the soft dipole mode leading to breakup.  An optical model approach is used in \cite{canto,taki, huss} while a coupled channels approach is used in \cite{dasso}.  In the optical model approach, the breakup channel is taken into account using a polarization potential while in the coupled channel calculations, breakup is treated like an additional channel with its coupling strength taken from the measured soft dipole excitation.  The coupled channels approaches naturally lead to enhanced cross sections as these couplings add to the cross section while the optical model approaches lead to reduced cross sections. In most  cases considered in Fig 1, the cross section resulting from a simple one-dimensional barrier penetration model is also shown.  It seems clear that a measurement of the fusion excitation function for the $^{11}$Li + $^{208}$Pb reaction would be valuable in resolving the differences between the various predictions shown in Fig. 1.
\subsection{Prior work}
The general problem of the near barrier fusion and breakup reactions of weakly bound nuclei has been studied, with differing conclusions.  For the $^{6}$He + $^{209}$Bi reaction \cite{deyoung, kolata}  enhanced sub-barrier fusion was observed while in the $^{6}$He + $^{238}$U reaction \cite{raabe}, a possible suppression of sub-barrier fusion was observed.  For the $^{9}$Be + $^{208}$Pb reaction \cite{dasgupta}, $^{9}$Be + $^{209}$Bi reaction \cite{cosmo1}, $^{6,7}$Li + $^{209}$Bi reactions \cite{dasgupta}, and $^{6}$Li + $^{208}$Pb reaction \cite{wu}, a large suppression of fusion above the barrier has been observed. For the $^{8}$Li + $^{208}$Pb reaction, a suppression of fusion at above barrier energies was observed \cite{agui}, with the $^{8}$Li projectile breaking up to give $^{7}$Li which fused with $^{208}$Pb.   For the $^{11}$Be + $^{209}$Bi  \cite{yoshida, cosmo2} and the $^{19}$F + $^{208}$Pb systems \cite{ernst}, the effect of breakup on the fusion cross section was negligible.  Recent review articles dealing with the general subject of the fusion of weakly bound nuclei are available \cite {felix, cgdh, keeley}

For the $^{11}$Li + $^{208}$Pb reaction, more recent theoretical work \cite{canto2} suggests that incomplete fusion and sequential complete fusion are negligible processes.  Additional theoretical treatments of the fusion of $^{11}$Li with  $^{208}$Pb have been made recently \cite{duhan,adel}.  Elastic scattering measurements for $^{9,11}$Li + $^{208}$Pb have been performed \cite{cubero}. A general universal framework for analyzing fusion excitation functions for weakly bound nuclei has been suggested \cite{canto3,canto4,gomes1,gomes2}.  The role of neutron transfer in fusion reactions with weakly bound nuclei has been studied recently \cite{sarg}

Our group has been engaged in a deliberate careful approach to measuring the $^{11}$Li + $^{208}$Pb fusion excitation function.  We started by studying the fusion of $^{9}$Li with $^{70}$Zn at ISAC at TRIUMF. $^{70}$Zn was chosen as the target nucleus because the ``energy limit" (at that time) of the ISAC beams of ~ 1.7 A MeV prevented one reaching the fusion barrier in heavier systems. The results of this study \cite{wdl1} showed a large sub-barrier fusion enhancement for the reaction of $^{9}$Li with $^{70}$Zn that was not accounted for by current models of fusion. Attempts to describe these results \cite{zaggy,bala} required unusual mechanisms to enhance sub-barrier fusion in these systems.  Zagrebaev et al. \cite{zaggy} found that standard coupled channels calculations along with neutron transfer were not able to describe the observed sub-barrier fusion and postulated ``di-neutron transfer" to account for the observed data.  Balantekin and Kocak \cite{bala} also found that coupled channels calculations including inelastic excitation and one-neutron transfer failed  to reproduce the data and suggested the possible formation of a molecular bond accompanied by two-neutron transfer to account for the observed behavior.  In this approach, the neutron-rich $^{70}$Zn contributes two neutrons to form the $^{11}$Li halo structure in the nuclei at contact, which enhances the fusion cross section.  The data \cite{wdl1} are well represented by this model.

We then measured the fusion excitation function for the $^{9}$Li + $^{208}$Pb reaction for near barrier projectile c.m. energies of 23.9 to 43.0 MeV using the ISAC2 facility at TRIUMF \cite{amv}.  The $\alpha$-emitting evaporation residues ($^{211-214}$At) were stopped in the $^{208}$Pb target and their decay was measured.  The At yields at each energy were in good agreement with the predictions of statistical model codes \cite{willi,matt-willi,zaggyweb} (Fig. 2).

The statistical mode codes are based on evaluating the terms in the general equation for the production of an evaporation residue, $\sigma$$_{EVR}$, as 
\setcounter{equation}{0}
\begin{equation}
\sigma _{EVR}=\sum_{J=0}^{J_{\max }}\sigma
_{capture}(E_{c.m.},J)P_{CN}(E^{*},J) W_{sur}(E^{*},J)
\end{equation}
where $\sigma _{capture}(E_{c.m.},J)$ is the capture cross section at center of mass energy E$_{c.m.}$ and spin J. P$_{CN}$ is the probability that the projectile-target system will evolve from the contact configuration  inside the
fission saddle point to form a completely fused system rather than
re-separating (quasifission, fast fission). W$_{sur}$ is
the probability that the completely fused system will de-excite by neutron emission rather than fission.  For fusion studies involving weakly bound nuclei, it is probably appropriate to use the relation for P$_{CN}$ as 
\begin{equation}
P_{CN}(E^{*},J)=1-P_{BU}(E^{*},J)
\end{equation} 
where P$_{BU}$ refers to the probability that the projectile broke up rather than fused.   In both statistical model calculations, the breakup probability was assumed to be zero, i.e., P$_{CN}$ was assumed to be 1.

 For the HIVAP calculations \cite{willi} shown in Fig. 2, , the ``Reisdorf-Sch$\ddot{a}$del" parameters \cite{matt-willi} were used.  For the calculations labeled ``Zagrebaev", the Nuclear Reactions Video Project applets \cite{zaggyweb} were used. For the latter approach, the capture cross section was calculated using the coupled channels method with inelastic excitations of the projectile and target nucleus being used.    The survival probability W$_{sur}$ can be written as 
\begin{equation}
W_{sur}=P_{xn}(E_{CN}^{\ast })\prod\limits_{i=1}^{i_{\max }=x}\left( \frac{%
\Gamma _{n}}{\Gamma _{n}+\Gamma _{f}}\right) _{i,E^{\ast }}
\end{equation}
where the index i is equal to the number of emitted neutrons and P$_{xn}$ is
the probability of emitting exactly x neutrons \cite{jackson}. In evaluating the excitation energy in  equation (3), we start at the excitation energy E* of the completely fused system and reduce it for each evaporation step by the binding energy of the emitted neutron and an assumed neutron kinetic energy of 2T where T (=(E*/a)$^{1/2}$) is the temperature of the emitting system.   For
calculating $\Gamma _{n}/\Gamma _{f}$, we have used the classical formalism
from Vandenbosch and Huizenga \cite{vh}
\begin{equation}
\frac{\Gamma _{n}}{\Gamma _{f}}=\frac{4A^{2/3}\left( E^{\ast }-B_{n}\right) 
}{k\left[ 2a^{1/2}\left( E^{\ast }-B_{f}\right) ^{1/2}-1\right] }\exp \left[
2a^{1/2}\left( E^{\ast }-B_{n}\right) ^{1/2}-2a^{1/2}\left( E^{\ast
}-B_{f}\right) ^{1/2}\right] 
\end{equation}
The constants k and a are taken to \ be 9.8 MeV and (A/12) MeV$^{-1}$, respectively. \ The
fission barriers B$_{f}$ are written as the sum of liquid drop, B$_{f}^{LD}$,
and shell correction terms as
\begin{equation}
B_{f}(E_{CN}^{\ast })=B_{f}^{LD}+U_{shell}
\end{equation}
where the shell correction energies , U$_{shell}$, to the LDM barriers are
taken from \cite{peter} , and the liquid drop barriers are taken from \cite{bill}.  Neutron binding energies, B$_{n}$ are taken from \cite{peter}. The fade-out of the shell corrections with increasing excitation energy is treated through the level density parameter using the method of  Ignatyuk et al. \cite{iggy} as
\begin{equation}
a=\widetilde{a}\left[ 1+\delta E\frac{1-\exp (-\gamma E)}{E}\right] 
\end{equation}
\begin{equation}
\widetilde{a}=0.073A+0.095B_{s}(\beta _{2})A^{2/3}
\end {equation}
where the shell damping parameter is taken to be 0.061.  
Collective enhancement effects of the level density are important for both deformed and spherical nuclei as are their dependence on excitation energy. \cite{sven,jung}.  We use the formalism of ref. \cite{zaggyweb} to express these effects via the equations
\begin{equation}
K_{coll}=K_{rot}(E)\varphi (\beta _{2})+E_{vib}(E)\cdot (1-\varphi(\beta _{2}))
\end{equation}
\begin{equation}
\varphi (\beta _{2})=\left[ 1+\exp \left( \frac{\beta _{2}^{0}-\left\vert\beta _{2}\right\vert }{\Delta \beta _{2}}\right) \right] ^{-1}
\end{equation}
\begin{equation}
K_{rot(vib)}(E)=\frac{K_{rot(vib)}-1}{1+\left[ \left( E-E_{\alpha }\right)/\Delta E_{\alpha }\right] }+1
\end{equation}
\begin{equation}
K_{rot}=\frac{J_{\bot }T}{\hbar ^{2}}
\end{equation}
\begin{equation}
K_{vib}=\exp (0.0555A^{2/3}T^{4/3})
\end{equation}

The calculated fusion-fission cross sections for the $^{9}$Li + $^{208}$Pb reaction (E$_{lab}$ = 24.8 - 44.9 MeV) range from 0 - 5 mb for the HIVAP calculations and from 0 - 180 mb for the methods of \cite{zaggyweb}.  Similarly the calculated fusion-fission cross sections for the $^{11}$Li + $^{208}$Pb reaction (E$_{lab}$ = 28.6 - 39.9 MeV) range from 0 - 3 mb for the HIVAP calculations and from 0 - 330 mb for the methods of \cite{zaggyweb}.

  The measured fusion excitation function for the $^{9}$Li + $^{208}$Pb reaction (Fig.  3) showed evidence for substantial sub-barrier fusion enhancement not predicted by current theoretical models or coupled channel calculations.  There was a suppression of the above barrier cross sections relative to these model predictions. 

We believe these observations are significant because $^{9}$Li is the ``core" of the two-neutron halo nucleus $^{11}$Li. Many calculations have suggested that in the interaction of $^{11}$Li with $^{208}$Pb, the $^{11}$Li will break up into two neutrons and the $^{9}$Li core, which,  in turn,  will fuse with the $^{208}$Pb nucleus.  In the study of Petrascu et al. \cite{petrascu} of the fusion of $^{9,11}$Li with Si at 11.2-15.2 A MeV, they found evidence that the $^{9}$Li fused with the Si, but in the case of $^{11}$Li there was emission of one or two neutrons prior to fusion.

In section II of this paper, we describe the experimental arrangements while in section III, we describe and discuss the results of the measurement.

\section {Experimental Methods}
\subsection{Setup and design}

The measurement of the fusion cross section for the $^{11}$Li + $^{208}$Pb reaction was carried out at the ISAC2 facility at TRIUMF.  Proton beams (500 MeV) with an intensity of $\sim$ 70$\mu$A struck Ta metal production targets.  Beams of radioactive $^{11}$Li were extracted with energies up to 18.4 keV, mass-separated by passage through two dipole magnets and accelerated to their final energy by radio frequency quadrupole and drift tube linear accelerators.  The details of the production of these secondary beams are discussed elsewhere \cite{pierre, marik}. The stable $^{7}$Li beam used to calibrate the efficiency of the experimental setup (see below) was generated using a local ion source.

A $^{11}$Li beam (40 MeV) impinged on a set of four $^{208}$Pb foils, backed by 0.54 mg/cm$^{2}$ Al foils.  The $^{208}$Pb target thicknesses were 0.859, 0.414, 0.605 and 1.019 mg/cm$^{2}$.  The $^{208}$Pb material was 99.00 $\%$ $^{208}$Pb, 0.70 $\%$ $^{207}$Pb and 0.30$\%$ $^{206}$Pb.  A schematic diagram of the experimental apparatus is shown in Figure 4.   The target/degrader foil assemblies were tilted at 45$^{\circ}$ with respect to the incident beam direction.  Each Pb target/degrader assembly was at the center of a cubical vacuum chamber, where four 300 mm$^{2}$ Canberra PIPS silicon detectors viewed the target/degrader assembly.   Photographs of the ``cubes" and their innards are shown in Figure 5.  The ``center of target" $^{11}$Li beam energies were 39.9, 36.5, 32.7, and 28.6 MeV in the four ``cubes", i.e., spanning c.m. energies of 37.9 to 27.1 MeV, from above to below the nominal interaction barrier.   The $^{11}$Li beam was pulsed on for $\leq$5 ns and shutoff for 172 ns, during which time, the $\alpha$-decay of any stopped evaporation residue was measured.  The $^{11}$Li  beam intensity was monitored by a Si detector mounted in the beam line behind the ``cubes".  The average $^{11}$Li beam intensity was $\sim$1250 p/s for the $\sim$34 hours the beam was on target (during the  5 day experimental period).

\subsection {Alpha decay measurements}

The fusion-like $\alpha$-emitting EVRs in the $^{11}$Li + $^{208}$Pb reaction are expected to be astatine isotopes.  (As in our studies of the $^{9}$Li + $^{208}$Pb reaction, we see no evidence for the formation of Pb, Bi or Po isotopes ($\sigma$$_{upper}$ $\sim$ 6 mb), presumably indicating that these incomplete fusion products were formed with smaller or negligible yields.) In Table 1, we summarize the decay properties of the At isotopes.  The measured $\alpha$-particle detector resolution of our cubical detector arrays with their thick target/degrader assemblies was $\sim$ 145 keV (FWHM).  If we take into account the observed tendency in these reactions \cite{dasgupta} to preferentially populate the high spin member of an isomeric pair, then it is a straightforward exercise to show that we can resolve the decays of $^{213}$At and $^{214}$At$^{m}$, but it is difficult to resolve the $\alpha$-particles emitted by $^{216}$At$^{m}$, $^{215}$At and $^{212}$At$^{m}$ on the basis of $\alpha$-particle energy. A typical alpha spectrum demonstrating this idea is shown in Figure 6.  (See below for another approach).

 The detected activities are produced and decay during irradiation in accord with the equations of radioactive decay.   All decays of the metastable states to lower lying states by IT decay are negligible.  For nuclei that are produced directly during the irradiation, the number of atoms present, N$_{2}$, after a ``beam on" period of t sec is given as 
\begin{equation}
N_{2}(t)=N_{2}(0)\exp (-\lambda _{2}t)+\frac{R_{2}}{\lambda _{2}}(1-\exp (-\lambda _{2}t))
\end{equation}
where N$_{2}$(0) is the number of nuclei present at the beginning of the period, R$_{2}$ is the rate of production ($\equiv$ N$_{target}$$\sigma$$\phi$), $\lambda$$_{2}$ the decay constant, N$_{target}$ the number of target atoms, $\sigma$ the cross section and $\phi$ the beam intensity.  During the ``beam off" period, the number of atoms decreases due to decay
\begin{equation}
N_{2}(t)=N_{2}(0)\exp (-\lambda _{2}t)
\end{equation}
It is straightforward to show that when the total ``beam on" time is long compared to the half-lives of the nuclide involved, the number of decays of product atoms per ``beam off" period is a constant fraction of the term $\frac{R_{2}}{\lambda _{2}}$.   Standard equations of production and decay were used to describe this decay which was detected after the end of each irradiation.

\subsection{Efficiency calibration}

To check that we understood all aspects of the measurement of nuclidic activities we also measured the yield of the evaporation residues $^{212,213}$Rn formed in the reaction of 34.90 MeV $^{7}$Li with $^{209}$Bi and compared our results to the previous measurement of Dasgupta, et al. \cite{dasgupta}.  In this calibration reaction, a single detector ``cube" assembly was used.  The $^{209}$Bi target thickness was 0.477 mg/cm$^{2}$ and  the $^{7}$Li center-of-target beam energy was 34.90 MeV.  The $^{7}$Li$^{3+}$ beam intensity was 130 picoamperes.  A geometry factor for detecting $\alpha$-particles in a single ``cube" of about 20$\%$ was used in the calculation. Our results are shown in Table II and Figure 7.  The agreement  between our results and those of \cite{dasgupta} is acceptable,  indicating we are able to reproduce known information about similar reactions.

\section{Results and Discussion}

\subsection{Cross Sections--Comparison with statistical model calculations}

As discussed above, the identification of $^{213}$At and $^{214}$At is straightforward as is the calculation of the observed nuclidic production cross sections.  These cross sections are tabulated in Table III.  There is an ambiguity in the identification of $^{212}$At$^{m}$, $^{215}$At and $^{216}$At$^{m}$ based upon alpha spectroscopy alone.  However, we note that in our study of the interaction of $^{9}$Li with $^{208}$Pb, statistical models were successful in describing the observed nuclidic At production cross sections and we turn to them again.  In Figure 8, we show the predictions of the HIVAP and Zagrebaev models for the fusion of $^{11}$Li with $^{208}$Pb.  Unfortunately, there are disagreements between the model predictions for some radionuclides.  We do note however that both models predict a very small $^{216}$At production cross section.  We shall assume that this nuclide, which would be a complete fusion product, has a negligibly small production cross section and remove it from the $^{212}$At$^{m}$/$^{215}$At ambiguity.

Because of the differences between the statistical model predictions and the limited resolution for decay $\alpha$-particles due to the experimental geometry, we shall tabulate (Table III) a cross section value representing the sum of the $^{212}$At$^{m}$ and $^{215}$At cross sections.  When comparing these data to the statistical model calculations, we shall also sum the values of the predicted cross sections for $^{212}$At and $^{215}$At.  

One possibility that we need to consider is that the evaporation residues result from the breakup of the $^{11}$Li projectile followed by the fusion of the $^{9}$Li core with $^{208}$Pb.  (The energetics of some possible breakup processes are shown in Table IV.  Clearly the breakup of $^{11}$Li into $^{9}$Li with the subsequent fusion of the $^{9}$Li fragment with $^{208}$Pb is energetically possible.  The other breakup transfer reactions in Table IV were not seen in the EVR yields although the `breakup-two neutron capture" process leads to a radionuclide, $^{210}$Po, whose half-life is too long to be observed in this study.)

In Figures 9 and 10 we compare the observed At nuclidic yields with (a) the predicted cross sections (HIVAP, Zagrebaev) for the complete fusion of $^{11}$Li with $^{208}$Pb (Fig. 9) (b) the observed At yields for the $^{9}$Li + $^{208}$Pb reaction, representing the outcome of incomplete fusion (Fig. 10).  (In the case of $^{214}$At, there is limited data for the $^{9}$Li + $^{208}$Pb reaction, so we have compared the yields with statistical model predictions.)  In making the incomplete fusion comparison, we have shifted the c.m. energies of the $^{11}$Li beam by 9/11.

The observed $^{212/215}$At yields are in reasonable agreement with the statistical model predictions for complete fusion, especially the calculations using the Zagrebaev model.  The shifted $^{212/215}$At yields are not very similar to the  $^{212/215}$At yields from the $^{9}$Li + $^{208}$Pb reaction.  One cannot absolutely rule out the possibility that the  $^{212/215}$At yields from the $^{11}$Li + $^{208}$Pb reaction are consistent with a breakup fusion process, but the excitation function would have a very long low energy tail.

The observed $^{213}$At yields are in rough agreement with the statistical model predictions for complete fusion , but one is impressed by the striking concordance between the shifted measured yields and the measured excitation function for the $^{9}$Li + $^{208}$Pb reaction.  It is hard to imagine this agreement is by chance.

The observed $^{214}$At yields are in rough agreement with the statistical model predictions for complete fusion.  Because there is very little data on the $^{214}$At yields in the $^{9}$Li + $^{208}$Pb reaction (Fig. 2), ,we have chosen to compare the shifted yields with the statistical model predictions for the $^{9}$Li + $^{208}$Pb reaction (Fig. 10).  

In all the statistical model calculations for complete fusion, we have assumed P$_{BU}$ = 0.  For $^{212,215}$At, that appears to be a good assumption (Fig. 9, 10).  For $^{213}$At, that assumption does not appear to be correct (Fig. 10).  The uncertainties in the measured data and the disagreement between the statistical model predictions are too large to support a detailed analysis, but one can note that the assumption that P$_{BU}$ $\sim$ 0.8 will produce a reasonable agreement between the measured data and the statistical model calculations for complete fusion.

We conclude from our analysis of the individual nuclidic yields that both complete fusion and breakup fusion are occurring in the interaction of $^{11}$Li with $^{208}$Pb.

\subsection{Comparison with theory}

If we { \bf arbitrarily} assign the yields of $^{213}$At to ``breakup processes" or ``incomplete fusion" and the yield of $^{212/215,214}$At as ``complete fusion", we can calculate (and tabulate (Table V) the ``total fusion" cross section $\sigma$$_{TF}$ as
\begin{equation}
\sigma_{TF} = \sigma_{ICF} + \sigma_{CF}
\end {equation}
.  In Fig. 11, we compare the various theoretical predictions for the fusion cross section in the $^{11}$Li + $^{208}$Pb reaction with our data for complete fusion (CF) and ``total" fusion (TF).  Apart from the lowest energy studied, we conclude that all  the calculations substantially overestimate the magnitude of the complete fusion (and/or total fusion) cross sections. At E$_{c.m.}$ = 27.2 MeV, the predictions of \cite{huss} are in good agreement with the measured data.
 
 To get some idea of the macroscopic parameters for the combined fusion/breakup interaction of $^{9}$Li with $^{208}$Pb, we focus our attention on the total fusion cross sections.  We use the coupled channels formalism described earlier \cite{zaggyweb} with the optical model parameters established by Cubero et al. \cite{cubero} that describe the elastic scattering of $^{9,11}$Li by $^{208}$Pb.  In this way, we are presenting a consistent picture of the interaction of $^{9,11}$Li with $^{208}$Pb.  We compare the predicted total interaction cross sections with the measured total fusion cross sections in Fig. 12.  The $^{9}$Li data seem to be adequately represented by the same optical model parameters used to describe the elastic scattering.  The total fusion cross-sections for the $^{11}$Li + $^{208}$Pb  reaction differ significantly from the coupled channels calculations.

 \subsection{Comparison with previous measurements}

In Figure 13, we show the measured fusion excitation functions for the $^ {6}$Li + $^{208}$Pb\cite{wu}, $^{7}$Li + $^{208}$Pb \cite{dasgupta} , $^{8}$Li + $^{208}$Pb \cite{agui},  the $^{9}$Li + $^{208}$Pb \cite{amv} and the $^{11}$Li + $^{208}$Pb reaction (this work).  What is presented in Figure 13 are the ``reduced" excitation functions in which each fusion cross section is divided by $\pi$R$_{B}$$^{2}$ and each energy is shown as E$_{c.m.}$/V$_{B}$ where R$_{B}$ and V$_{B}$ are the fusion radii and barrier heights in the semiempirical Bass model \cite{bass}.  All the ``reduced" excitation functions appear to be similar with the exception of the  $^{11}$Li + $^{208}$Pb reaction  .  This would indicate the basic differences between these different Li nuclei in their interaction with $^{208}$Pb are geometrical in origin except for the  $^{11}$Li + $^{208}$Pb reaction which is fundamentally different.

 \subsection{Future work}
 It is clear that the interesting and unexpectedly large breakup cross-section in the $^{11}$Li + $^{208}$Pb reaction should be investigated further. Some suggested extensions of this work are: 
 (a) Improvements in the  $^{11}$Li beam intensity and the on-target time of the beam to reduce the statistical uncertainties in the measured data and to extend the measurements to lower excitation energies where more direct comparison to theoretical predictions can be made.
 (b) With improvements in the total beam doses, more inclusive measurements of the evaporation residues and non-fusing breakup nuclei would be helpful.
 (c) Measurements of the interaction of $^{9,11}$Li with other target nuclei, such as $^{144,154}$Sm would be useful.
 
\section{Conclusions}

We conclude that:
(a) It is possible to measure the evaporation residue production cross-sections resulting from the interaction of  $^{11}$Li with $^{208}$Pb using current generation radioactive beam facilities.
(b) Both complete fusion and breakup fusion processes occur in the interaction of $^{11}$Li with $^{208}$Pb.
(c)An important breakup process leads to the fusion of the $^{9}$Li fragment with $^{208}$Pb .

\begin{acknowledgments}

We thank the operations staff of the cyclotron and ISAC, and Marco Marchetti and Robert Laxdal for providing the $^{7,11}$Li beams.

This work was supported, in part, by the Office of High Energy and Nuclear Physics, Nuclear Physics Division, U.S. Dept. of Energy, under Grant DE-FG06-97ER41026 and TRIUMF and the Natural Sciences and Engineering Research Council of Canada.

\end{acknowledgments}

\newpage
\begin{table}[tbp]
\caption{Decay properties of the astatine EVRs observed in this work}
\begin{ruledtabular}
\begin{tabular}{ccc}
Isotope & t$_{1/2}$ (s) & E$_{\alpha}$ (keV) (\% abundance) \\
$^{212}$At & 0.314 & 7679 (82);7616(16) \\
$^{212}$At$^{m}$ & 0.119 & 7837(66);7900(31.5) \\
$^{213}$At & 125x10$^{-9}$ & 9080(100) \\
$^{214}$At & 558x10$^{-9}$ & 8819(98.95) \\
$^{214}$At$^{m1}$ & 265x10$^{-9}$ & 8877(100)  \\
$^{214}$At$^{m2}$ & 760x10$^{-9}$ & 8782(99.18) \\
$^{215}$At & 0.10 x 10$^{-3}$ & 8026 (99.95) \\
$^{216}$At & 0.30 x 10$^{-3}$ & 7802 (97)  \\
$^{216}$At$^{m}$ & 0.1 x 10$^{-3}$ &  7960 (100) \\

\end{tabular}
\end{ruledtabular}
\end{table}

\newpage
\begin{table}[tbp]
\caption{Comparison of our EVR measurements for the $^{7}$Li + $^{209}$Bi reaction with ref. \cite{dasgupta}}
\begin{ruledtabular}
\begin{tabular}{ccc}
Isotope & Cross section (mb) \cite{dasgupta} & Cross section (mb)-this work \\
$^{213}$Rn & 195.0 $\pm$ 3.2 & 211.2 $\pm$ 8.1\\
$^{212}$Rn & 154.3 $\pm$ 4.9 & 158.2 $\pm$ 2.9 \\
Fission &  3.16 $\pm$ 0.03 &  Not measured \\
Complete Fusion Cross Section & 352.5 $\pm$ 5.9 & 372.6 $\pm$ 11.3 \\
\end{tabular}
\end{ruledtabular}
\end{table}

\newpage
\begin{table}[tbp]
\caption{Measured nuclidic cross sections (mb)  for the $^{11}$Li + $^{208}$Pb reaction}
\begin{ruledtabular}
\begin{tabular}{cccc}
E$^{lab}_{cot}$(MeV)  & $^{212}$At/$^{215}$At & $^{213}$At & $^{214}$At  \\
28.6 & 24 $\pm$ 12 & 6 $\pm$ 6 & 12 $\pm$ 8  \\
32.7 & 35 $\pm$ 16 & 14 $\pm$ 10 & 28 $\pm$ 14  \\ 
36.5 & 20 $\pm$ 14 & 90 $\pm$ 30 & 209 $\pm$ 46  \\ 
39.9 & 248 $\pm$ 61 & 146 $\pm$ 46 & 248 $\pm$ 60 \\ 
\end{tabular}
\end{ruledtabular}
\end{table}

\newpage
\begin{table}[tbp]
\caption{Possible breakup channels in the $^{11}$Li + $^{208}$Pb reaction}
\begin{ruledtabular}
\begin{tabular}{cc}
Reaction & Q (MeV) \\
$^{11}$Li $\rightarrow$ $^{9}$Li + 2 n & -0.30 \\
$^{11}$Li $\rightarrow$ $^{7}$Li + 4 n & -6.40 \\
$^{11}$Li $\rightarrow$ $^{10}$He + p & -15.3 \\
$^{11}$Li $\rightarrow$ $^{9}$He + p +  n & -15.5 \\
$^{11}$Li $\rightarrow$ $^{8}$He + 2 n  + p& -14.2 \\
$^{11}$Li  + $^{208}$Pb $\rightarrow$ $^{9}$Li + $^{210}$Pb& + 8.8 \\
$^{11}$Li  + $^{208}$Pb $\rightarrow$ $^{8}$Li + $^{211}$Pb& + 8.5 \\
$^{11}$Li  + $^{208}$Pb $\rightarrow$ $^{9}$Li + $^{212}$Pb& + 11.7 \\
$^{11}$Li  + $^{208}$Pb $\rightarrow$ 2n + $^{217}$At& -1.5 \\

\end{tabular}
\end{ruledtabular}
\end{table}

\newpage
\begin{table}[tbp]
\caption{Cross sections (mb)  for the $^{11}$Li + $^{208}$Pb reaction}
\begin{ruledtabular}
\begin{tabular}{cccc}
E$^{lab}_{cot}$(MeV)  & $\sigma$$_{CF}$ & $\sigma$$_{ICF}$ & $\sigma$$_{TF}$   \\
28.6 & 36 $\pm$ 14 &6 $\pm$ 6 & 42 $\pm$ 15 \\
32.7 & 63 $\pm$ 21 & 14 $\pm$ 10 & 77 $\pm$ 18 \\ 
36.5 & 229 $\pm$ 48 & 90 $\pm$ 30 & 319 $\pm$ 57 \\ 
39.9 & 496 $\pm$ 86 & 146 $\pm$ 46 & 642 $\pm$ 97 \\ 
\end{tabular}
\end{ruledtabular}
\end{table}

\newpage 
\begin{figure}[tbp]
\includegraphics [scale=0.60]{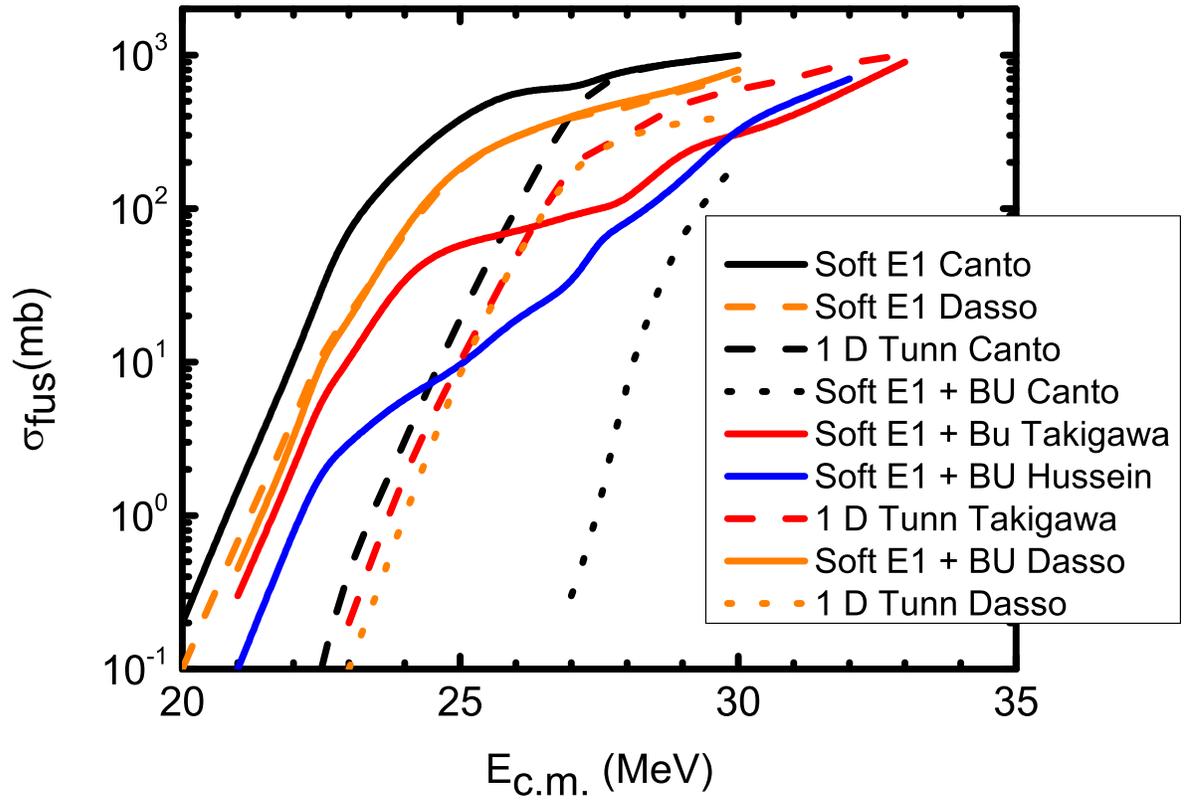}
\caption{(Color online)Various theoretical predictions for the $^{11}$Li + $^{208}$Pb fusion excitation function, after \cite{sig1,sig2}.  See text for a detailed discussion.}
\label{fig1}
\end{figure}

\begin{figure}[tbp]
\begin{center}
\epsfxsize 20.0cm
\epsfbox{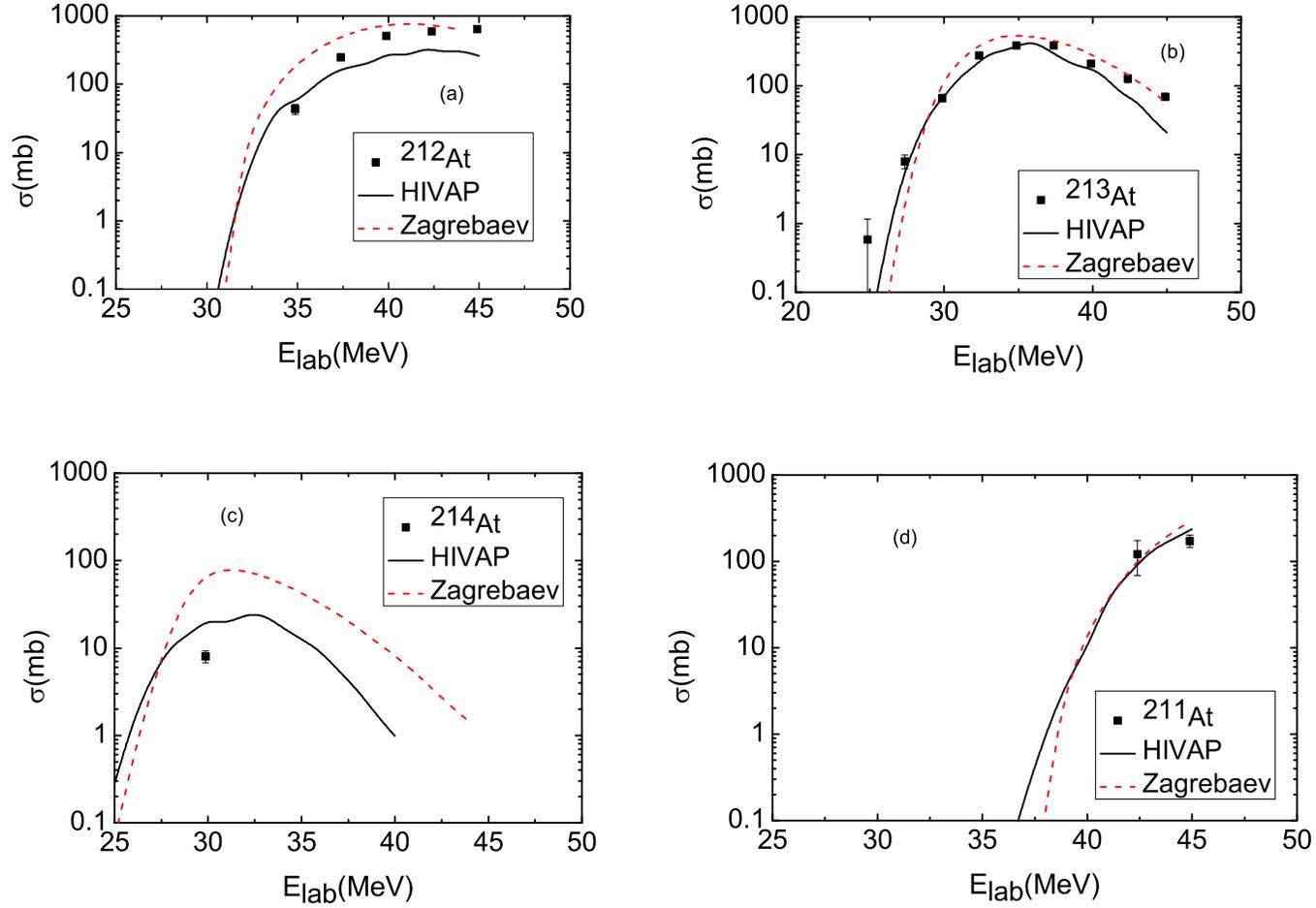}
\caption{(Color on-line) Comparison of measured nuclidic yields (data points) from the $^{9}$Li + $^{208}$Pb reaction with predictions of \cite{willi, matt-willi,zaggyweb} (lines).}
\end{center}
\end{figure}

\begin{figure}[tbp]
\epsfxsize 12.0cm
\epsfbox{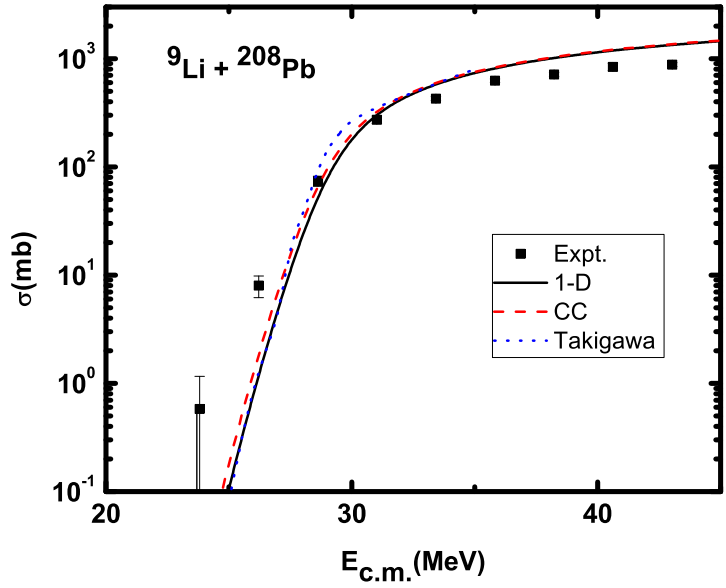}
\caption{(Color on-line) Comparison of the measured fusion excitation function \cite{amv} for the $^{9}$Li + $^{208}$Pb reaction with the predictions of coupled channels calculations and with \cite{takagawa}. }
\end{figure}

\newpage
\begin{figure}[tbp]
\epsfxsize 12.0cm
\epsfbox{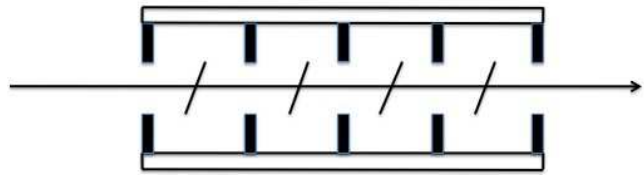}
\caption{ Schematic diagram of the experimental apparatus}
\end{figure}

\begin{figure}[h]
\begin{minipage}{30pc}
\begin{center}
\includegraphics[width=30pc]{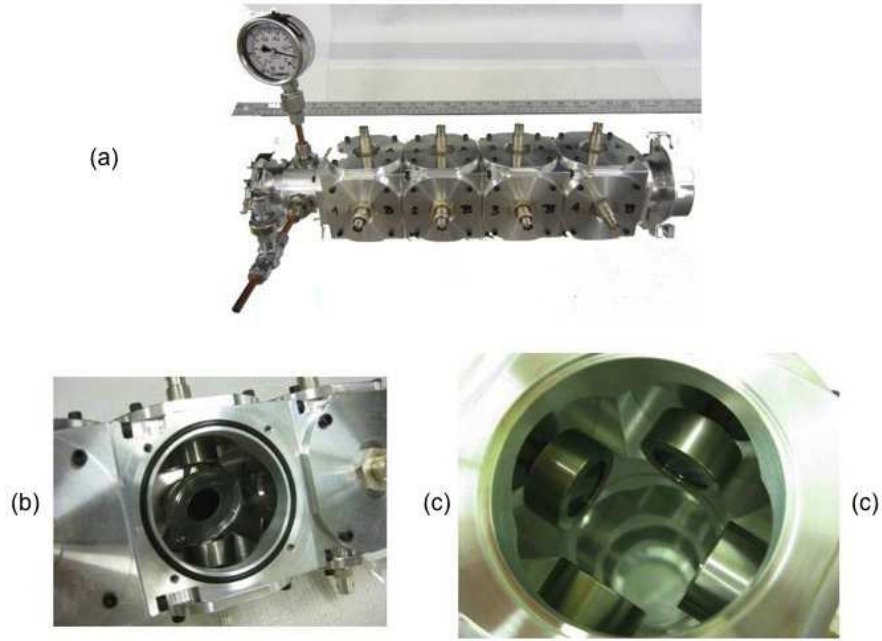}
\end{center}
\caption{(Color online) Photographs of cubical scattering chamber array showing (a) overall view,  (b) the target and detectors and (c)  the detectors only.}
\end{minipage} 
\end{figure}

\begin{figure}[h]
\begin{minipage}{40pc}
\begin{center}
\includegraphics[width=40pc]{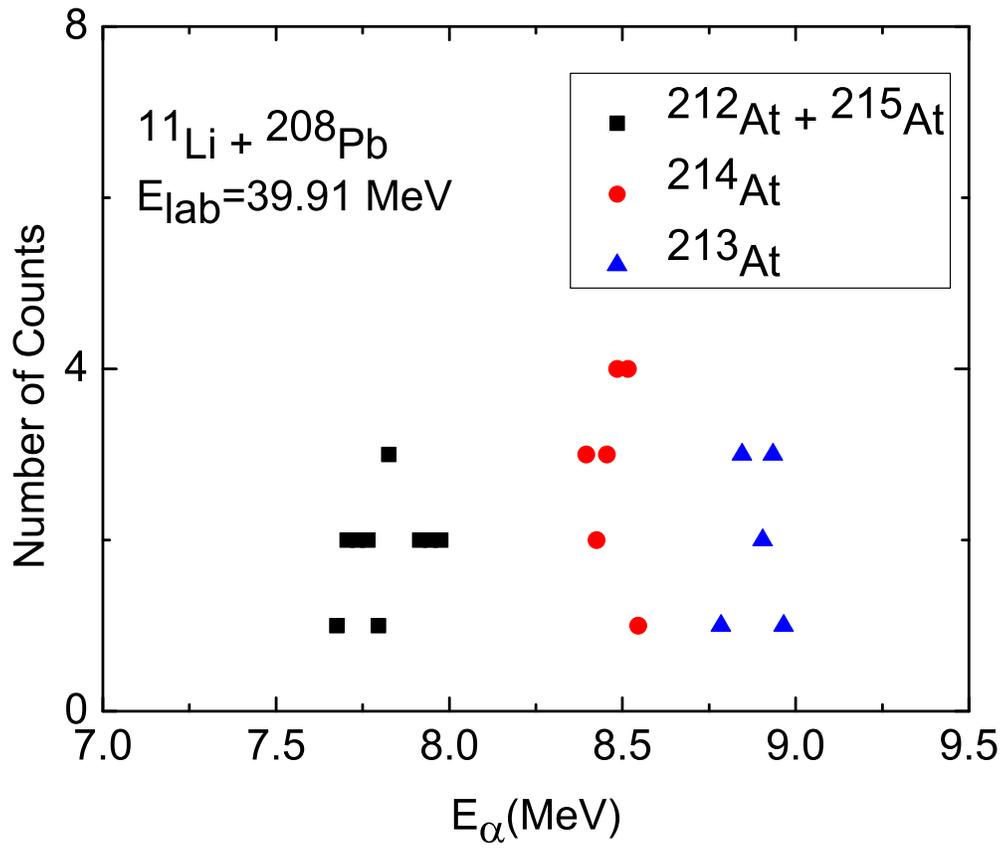}
\end{center}
\caption{(Color online) A typical alpha spectrum.}
\end{minipage} 
\end{figure}


\begin{figure}[tbp]
\epsfxsize 12.0cm
\epsfbox{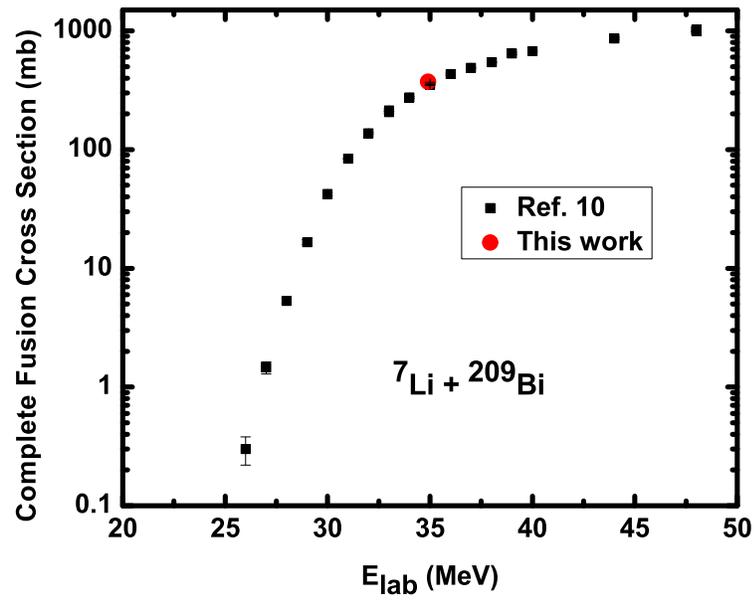}
\caption{(Color on-line) Comparison of EVR measurements for the $^{7}$Li + $^{209}$Bi reaction.}
\end{figure}

\begin{figure}[tbp]
\epsfxsize 20.0cm
\epsfbox{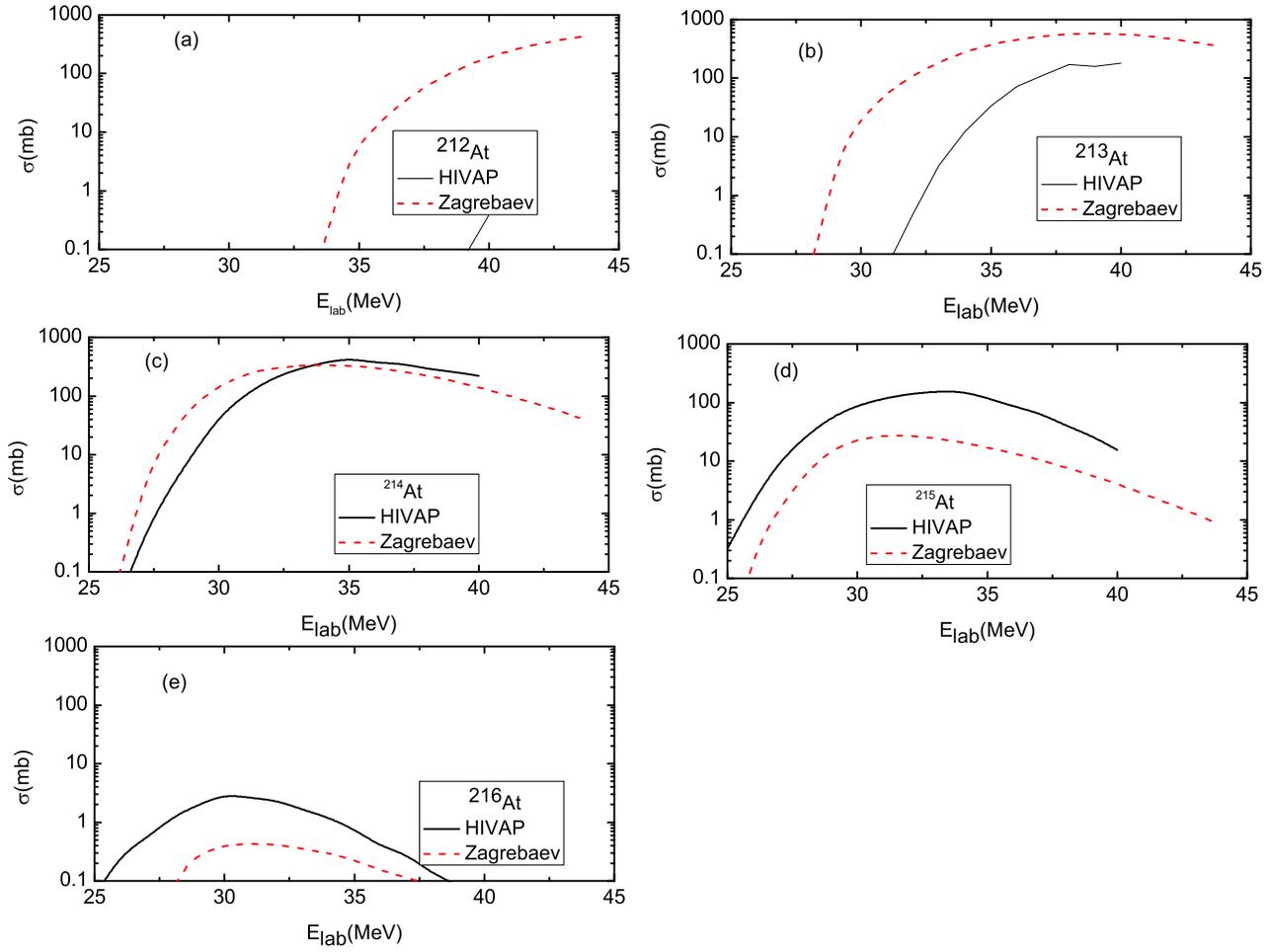}
\caption{(Color on-line) Comparison of predictions of \cite{willi, matt-willi,zaggyweb} (lines) for  the $^{11}$Li + $^{208}$Pb reaction.}
\end{figure}

\begin{figure}[tbp]
\epsfxsize 20.0cm
\epsfbox{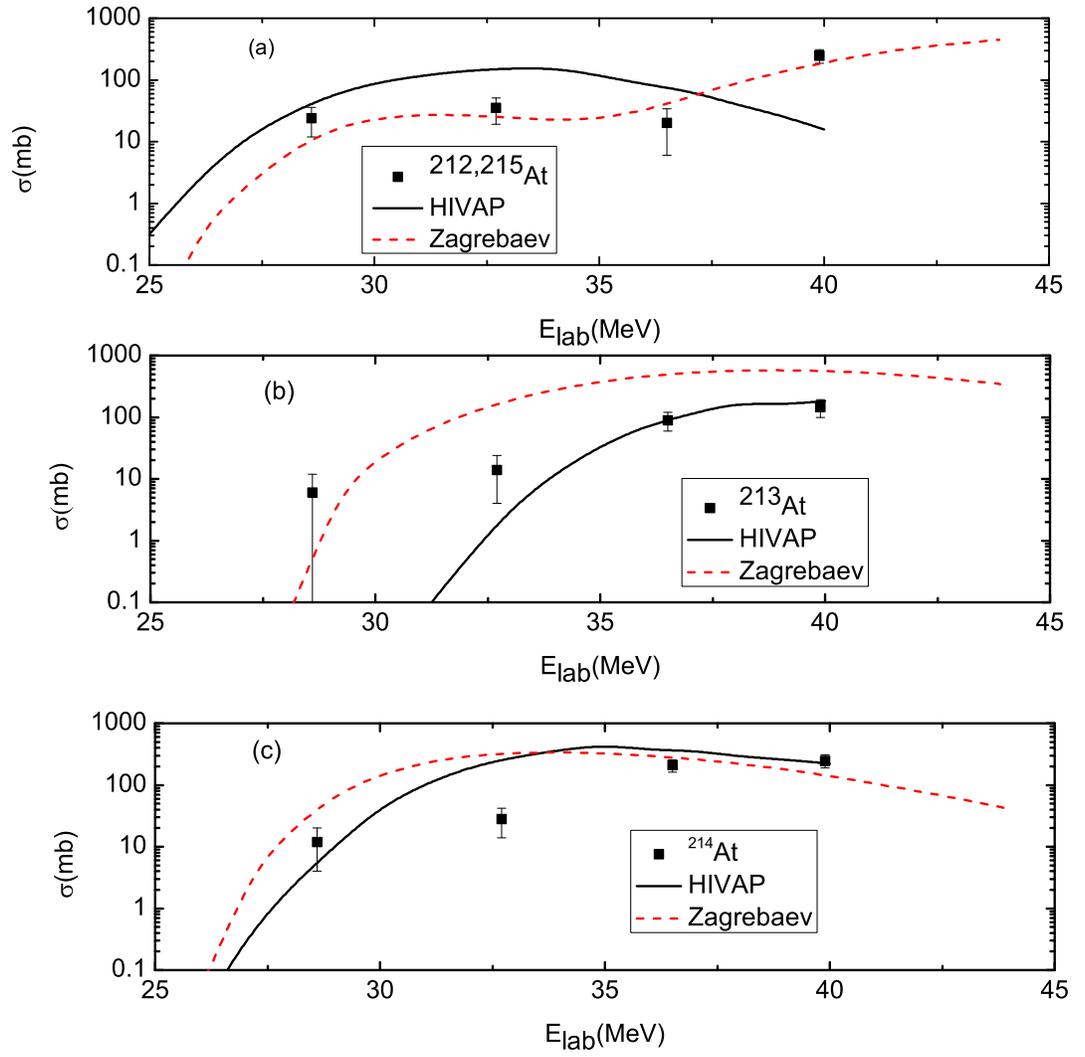}
\caption{(Color on-line) Comparison of predictions of \cite{willi, matt-willi,zaggyweb} (lines) for  complete fusion in the $^{11}$Li + $^{208}$Pb reaction with the measured data.}
\end{figure}

\begin{figure}[tbp]
\epsfxsize 15.0cm
\epsfbox{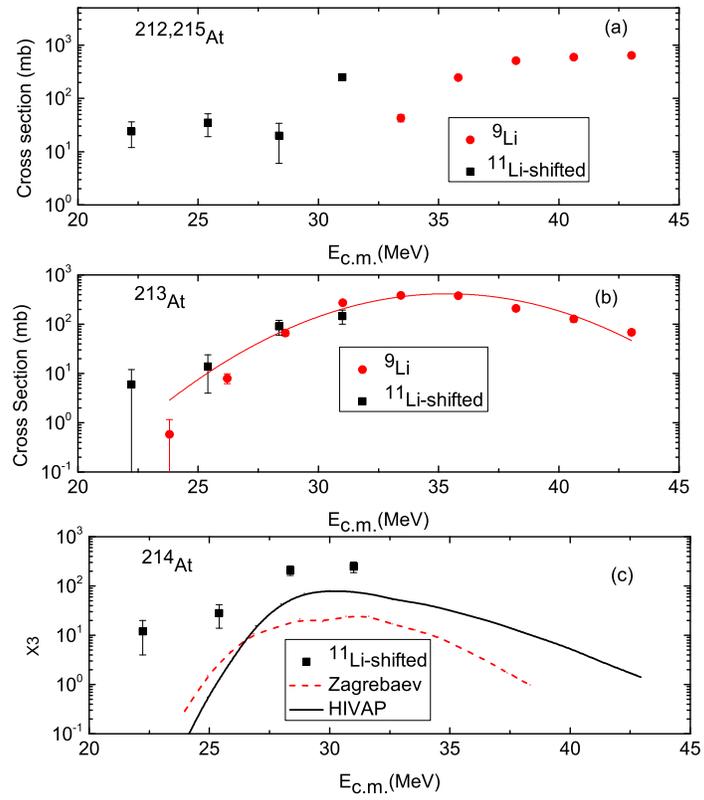}
\caption{(Color on-line) Comparison of measured nuclidic yields (data points) from the $^{9}$Li + $^{208}$Pb reaction with the shifted yields from the $^{11}$Li + $^{208}$Pb reaction.}
\end{figure}

\newpage 
\begin{figure}[tbp]
\includegraphics [scale=0.70]{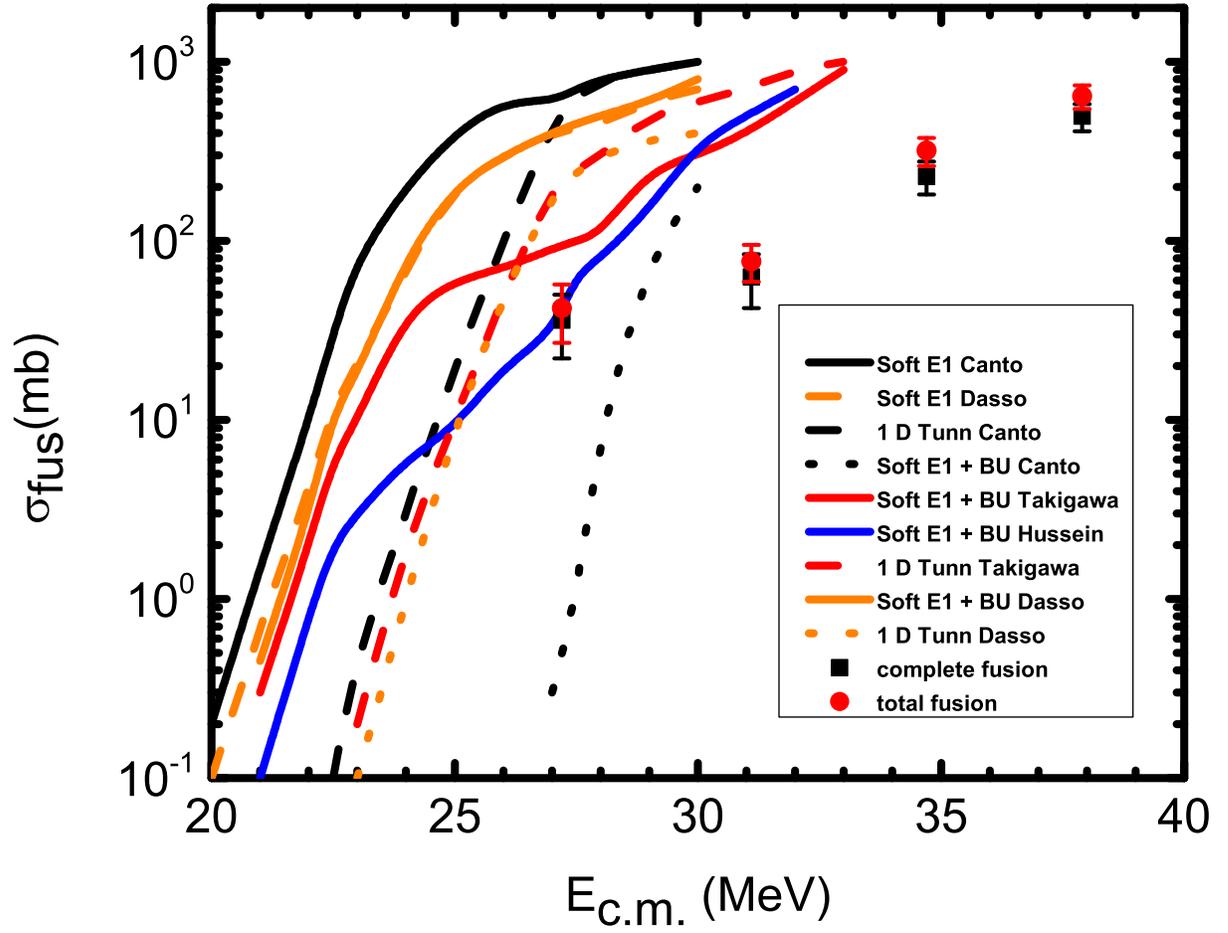}
\caption{(Color online)Comparison of various theoretical predictions for the $^{11}$Li + $^{208}$Pb fusion excitation function, after \cite{sig1,sig2},  with our data.}
\end{figure}

\newpage 
\begin{figure}[tbp]
\includegraphics [scale=0.70]{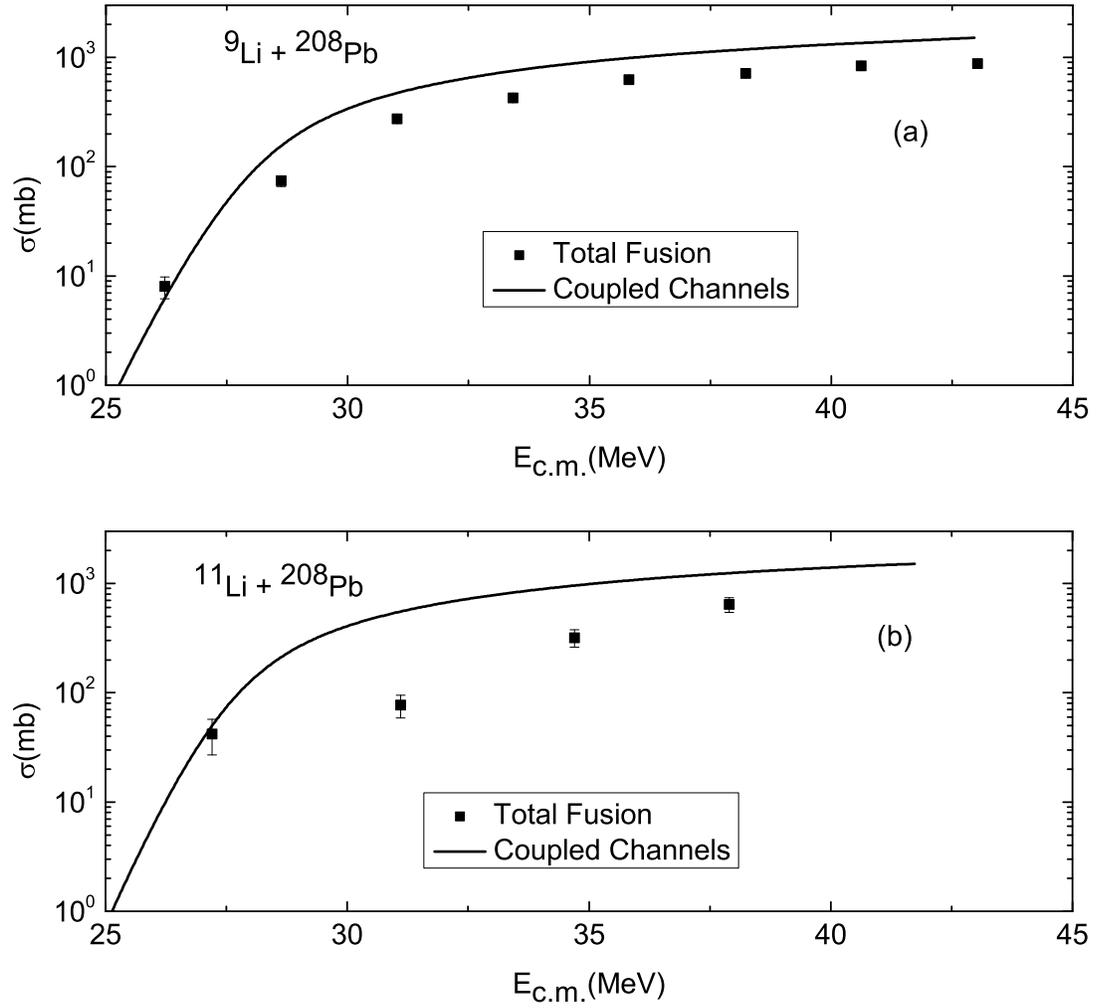}
\caption{Comparison of coupled channels calculations and data for the total interaction cross sections for $^{9,11}$Li with $^{208}$Pb.}
\end{figure}

\begin{figure}[tbp]
\epsfxsize 12.0cm
\epsfbox{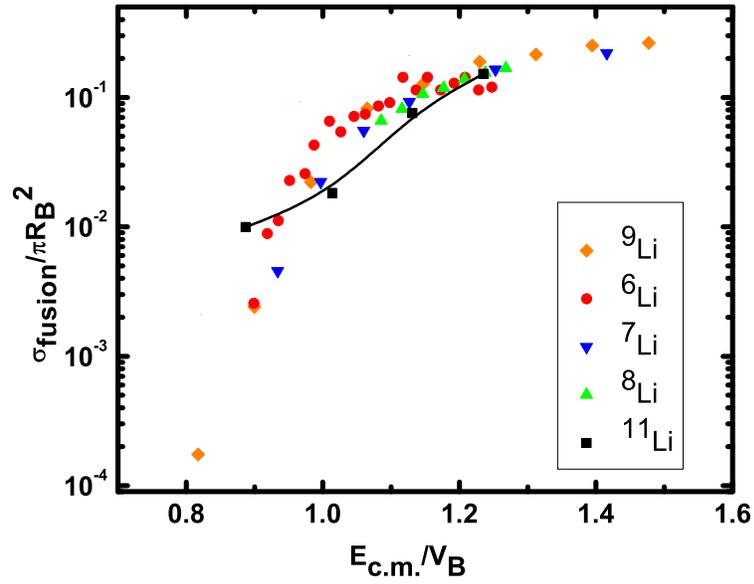}
\caption{(Color on-line) Comparison of the ``reduced" fusion excitation functions for the $^ {6}$Li + $^{208}$Pb\cite{wu}, $^{7}$Li + $^{208}$Pb \cite{dasgupta} , $^{8}$Li + $^{208}$Pb \cite{agui}, the $^{9}$Li + $^{208}$Pb reactions \cite{amv} and the $^{11}$Li + $^{208}$Pb reaction (this work). }
\end{figure}

\end{document}